\documentclass[sigconf]{acmart}
\AtBeginDocument{%
  \providecommand\BibTeX{{%
    \normalfont B\kern-0.5em{\scshape i\kern-0.25em b}\kern-0.8em\TeX}}}

\copyrightyear{2023}
\acmYear{2023}
\setcopyright{rightsretained}
\acmConference[CHI '23]{Proceedings of the 2023 CHI Conference on Human Factors in Computing Systems}{April 23--28, 2023}{Hamburg, Germany}
\acmBooktitle{Proceedings of the 2023 CHI Conference on Human Factors in Computing Systems (CHI '23), April 23--28, 2023, Hamburg, Germany}\acmDOI{10.1145/3544548.3581001}
\acmISBN{978-1-4503-9421-5/23/04}

\newcommand{\interviewprefix}{P}
\makeatletter
\newcommand{\interview}[1]{%
  \def\nextitem{\def\nextitem{, }}
  (\@for\el:=#1\do{\nextitem\interviewprefix\el})%
}
\newcommand{\longquote}[2]{\begin{quote}“\textit{#1}” -- P{#2}\end{quote}}%
\newcommand{\shortquote}[1]{“\textit{#1}”}%
\makeatother%

\usepackage{enumitem}

\begin{document}

\title[Understanding How Explainability Can Support Human-AI Interaction]{"Help Me Help the AI": Understanding How Explainability Can Support Human-AI Interaction}

\author{Sunnie S. Y. Kim}
\orcid{}
\affiliation{%
  \institution{Princeton University}
  \city{Princeton}
  \state{New Jersey}
  \country{USA}
}

\author{Elizabeth Anne Watkins}
\authornote{Most work was done during Postdoctoral Research Appointment at the Princeton Center for Information Technology Policy and Human-Computer Interaction Program.}
\affiliation{
  \institution{Intel Labs}
  \city{Santa Clara}
  \state{California}
  \country{USA}
}

\author{Olga Russakovsky}
\affiliation{%
  \institution{Princeton University}
  \city{Princeton}
  \state{New Jersey}
  \country{USA}
}

\author{Ruth Fong}
\affiliation{%
  \institution{Princeton University}
  \city{Princeton}
  \state{New Jersey}
  \country{USA}
}

\author{Andrés Monroy-Hernández}
\affiliation{%
  \institution{Princeton University}
  \city{Princeton}
  \state{New Jersey}
  \country{USA}
}

\renewcommand{\shortauthors}{Kim, Watkins, Russakovsky, Fong, Monroy-Hernández}

\begin{abstract}
Despite the proliferation of explainable AI (XAI) methods, little is understood about end-users' explainability needs and behaviors around XAI explanations. To address this gap and contribute to understanding how explainability can support human-AI interaction, we conducted a mixed-methods study with 20 end-users of a real-world AI application, the Merlin bird identification app, and inquired about their XAI needs, uses, and perceptions. We found that participants desire practically useful information that can improve their collaboration with the AI, more so than technical system details. Relatedly, participants intended to use XAI explanations for various purposes beyond understanding the AI's outputs: calibrating trust, improving their task skills, changing their behavior to supply better inputs to the AI, and giving constructive feedback to developers. Finally, among existing XAI approaches, participants preferred part-based explanations that resemble human reasoning and explanations. We discuss the implications of our findings and provide recommendations for future XAI design.
\end{abstract}




\maketitle

\section{Introduction}
\label{sec:intro}

Artificial Intelligence (AI) systems are ubiquitous: from unlocking our phones with face identification, to reducing traffic accidents with autonomous cars, to assisting radiologists with medical image analysis. Being able to better \emph{understand} these AI systems is becoming increasingly important---although what exactly that means is different in different settings: a smartphone user may want to \emph{understand} how best to position their face to quickly unlock their phone, a researcher may want to \emph{understand} what particular design decisions led to an autonomous car accident, and a radiologist may want to \emph{understand} where the medical decision support tool is looking in suggesting a particular diagnosis. 

Over the past years, numerous explainable AI (XAI) methods have been developed to provide transparency into these AI systems and make them more understandable to people (see \cite{arrieta2019explainable,fong2020thesis,gilpin2018explaining,Gunning_Aha_2019,samek2019book,Adadi2018survey,Das2020survey,Guidotti2018survey} for surveys).
However, arguably these are being developed without embracing the full spectrum of end-user needs. Particularly for computer vision AI systems (such as the ones described above), with millions of model parameters processing thousands of low-level image pixels, translating model outputs into understandable insights is so challenging that proposed XAI methods are frequently limited by what XAI researchers \emph{can do} rather than what AI end-users \emph{might need}. 

In this work, we connect XAI development with end-users and study a real-world context in which XAI methods might be deployed. Concretely, we set out to answer three research questions:
\begin{itemize}
    \item \textbf{RQ1}: What are end-users' XAI \textbf{needs} in real-world AI applications?
    \item \textbf{RQ2}: How do end-users \textbf{intend to use} XAI explanations\footnote{In this paper, we use the term ``XAI explanations'' to refer to explanations produced by XAI methods to explain specific AI system outputs.}? 
    \item \textbf{RQ3}: How are existing XAI approaches \textbf{perceived} by end-users?
\end{itemize}

In scoping our study, we focus on Merlin, an AI-based mobile phone application that uses computer vision to identify birds in user-uploaded photos and audio recordings. 
We chose Merlin because it is a widely-used application that allows us to connect with a diverse set of active end-users. Concretely, we conducted a mixed-methods study with 20 Merlin users who span the range from low-to-high AI background (representing both consumers and creators of AI systems) and low-to-high domain background (representing both users who know \emph{less} and \emph{more} about birding than the AI system). 

With each participant, we conducted an hour-long interview, which included a survey and an interactive feedback session, to understand their XAI needs, uses, and perceptions. Our study bridges the gap between XAI research done in the HCI and AI communities by directly connecting end-users of a real-world AI application with the XAI methods literature. We do so by mocking up four XAI approaches that could be potentially implemented into Merlin, i.e., heatmap, example, concept, and prototype-based explanations of the AI's outputs. 
The mock-up explanations enabled us to get concrete and detailed data about how participants intended to use XAI explanations, as well as how they perceived each approach, in an actual AI use context.

Through our study, we found:
\begin{itemize}[noitemsep,topsep=0pt]
    \item Participants' XAI needs varied depending on their domain/AI background and interest level. While participants were generally \emph{curious} about AI system details, those with high-AI background or notably high interest in birds had higher XAI needs. However, participants unanimously expressed a need for practically useful information that can improve their \emph{collaboration} with the AI, suggesting an important area of focus for future XAI development (RQ1, Sec. \ref{sec:xai_needs}).
    \item Participants intended to use XAI explanations for various purposes beyond understanding the AI's outputs: determining when to trust the AI, learning to perform the task better on their own without needing to consult the AI, changing their behavior to supply better inputs to the AI, and giving constructive feedback to the developers to improve the AI. This highlights the broad range of XAI needs that should be considered in XAI development (RQ2, Sec. \ref{sec:xai_use}).
    \item Among existing XAI approaches, participants preferred part-based explanations, i.e., concept~\cite{zhou2018ibd,ramaswamy2022elude} and prototype~\cite{chen2019protopnet,nauta2021prototree} based explanations. Participants found them similar to human reasoning and explanations, and  the most useful for the aforementioned purposes. This suggests that to the extent possible, the XAI community should pay particular attention to these methods,
    despite the challenges with their development and evaluation (RQ3, Sec. \ref{sec:xai_feedback}). 
\end{itemize}

Following our findings, we discuss XAI's potential as a medium for enhancing human-AI collaboration, and conclude with a set of recommendations for future XAI design. However, as with any case study, our findings and recommendations may have limited generalizability. This is an intentional trade-off made to gain an in-depth understanding of end-users' XAI needs, uses, and perceptions in a real-world context, in line with growing calls for human-centered XAI research~\cite{Ehsan2020HCXAI,Liao2021HCXAI,Ehsan2021HCXAI,Ehsan2022HCXAI,Liao2022Evaluation}. We are hopeful that our study design and insights will aid future XAI research in other contexts.

\section{Related work}
\label{sec:relatedwork}

\subsection{From algorithm-centered to human-centered XAI}

With the growing adoption of AI, there has been a surge of interest in explainable AI (XAI) research that aims to make AI systems more understandable to people.
XAI is one of the fastest growing fields with hundreds of new papers published each year. 
See~\cite{arrieta2019explainable,fong2020thesis,gilpin2018explaining,Gunning_Aha_2019,samek2019book,Adadi2018survey,Das2020survey,Guidotti2018survey,molnar2022,doshivelez2017rigorous,Abdul2018HCIsurvey,RudinEtAlSurvey2022,Sanneman2020SAFramework,Smuha2019EU} for in-depth surveys, and the following for examples of XAI research done in different disciplines: AI~\cite{Pedreschi2019AAAI,koh2020concept,fong2019extremal,kim2018tcav,Slack_TalkToModel_Understanding_Machine_2022}, HCI~\cite{Wang2019CHI,Hoffman2018,SmithRenner2020CHI,Zhang2022CHI}, social and cognitive science~\cite{Miller2017SocialScience,Miller2017,Confalonieri2019CogSci,Srinivasan2020CogSci,Taylor2018psychology,Broniatowski2021Psychology}, and philosophy~\cite{Mittelstadt2019FAccT,Kasirzadeh2021Philosophy,Baum2022Philosophy}. XAI is also increasingly being researched and applied in various domains, including but not limited to healthcare~\cite{Antoniadi2021Clinical,Lotsch2022Biomedicine,Smith2021ClinicalAI,Yang2022MedicalXAI,Zhang2022Diagnostics,Singh2020Imaging,Markus2021Healthcare,Poceviciute2020Pathology}, autonomous driving~\cite{Meteier2019AutomatedDriving,Atakishiyev2021autonomous,Omeiza2022autonomous}, energy and power systems~\cite{Machlev2022Energy}, and climate science~\cite{Mamalakis2022Climate}.

Much of the field's efforts originally focused on the \textit{algorithms}, i.e., on providing explanations of AI systems' inner workings and outputs,
rather than the \textit{people} or the \textit{context} where these systems are deployed. 
Recently, there has been a growing recognition that XAI methods cannot be developed ``in a vacuum'' without an understanding of people's needs in specific contexts~\cite{Ehsan2020HCXAI,Liao2021HCXAI,Ehsan2021HCXAI,Ehsan2022HCXAI,Liao_CHI_2020,Liao2022Evaluation}. 
In response, researchers have proposed \textit{conceptual frameworks} to characterize XAI needs based on people's roles~\cite{Preece2018Stakeholders,Tomsett2018roles,Langer2021stakeholders}, expertise~\cite{Mohseni2018survey}, or more fine-grained axes of knowledge and objectives~\cite{Suresh2021Stakeholders}.
Others interviewed \textit{industry practitioners} who work on AI products to identify their common XAI needs~\cite{Hong2020CSCW,Bhatt2020FAccT,Liao_CHI_2020}.

We join this relatively new line of research, called ``human-centered XAI''~\cite{Ehsan2020HCXAI,Liao2021HCXAI,Ehsan2021HCXAI,Ehsan2022HCXAI,Liao_CHI_2020,Liao2022Evaluation}, and foreground the people who use AI systems and their needs, goals, and contexts in \textit{understanding how explainability can support human-AI interaction}. In doing so, we build on the aforementioned frameworks to study end-users' explainability needs. Concretely, we developed a survey based on Liao and colleagues' XAI Question Bank~\cite{Liao_CHI_2020} to collect concrete data on which aspects of AI end-users want to know about.

\subsection{Understanding end-users' XAI needs}

Although human-centered XAI is an actively growing area of research, much of the work still focuses on developers rather than end-users of AI systems~\cite{Hong2020CSCW,Bhatt2020FAccT,Liao_CHI_2020}. 
This gap is unsurprising, since XAI methods have been primarily developed for and used by developers to inspect AI systems~\cite{Miller2017,Bhatt2020FAccT}.
But it is critical because end-users may have different explainability needs that XAI methods should but don't yet support.

Recently, some researchers began looking at end-users' XAI needs in context of specific applications~\cite{Tonekaboni2019clinicians,Cai2019CSCW,Cai2019CHI}.
Tonekaboni and colleagues~\cite{Tonekaboni2019clinicians} placed clinicians in \textit{hypothetical scenarios} where AI models are used for health risk assessment, and found that clinicians wanted to know what features the model uses so they can understand and rationalize the model's outputs.
In a lab setting, Cai and colleagues~\cite{Cai2019CSCW} studied clinicians' needs in their interaction with a \textit{prototype} AI model that can assist with cancer diagnoses, and found that clinicians desired overall information about the model (e.g., capabilities and limitations, design objective) in addition to explanations of the model's individual outputs.
In another lab setting, Cai and colleagues~\cite{Cai2019CHI} examined what needs pathologists have when using a \textit{prototype} AI model for retrieving similar medical images. They also studied how pathologists use their proposed refinement tools, finding that pathologists often re-purposed them to test and understand the underlying search algorithm and to disambiguate AI errors from their own errors.

These studies delivered rich insights. However, they studied \textit{hypothetical} or \textit{prototype} AI applications. Hence, an important question remains, which we tackle in this work: \textit{What are end-users' XAI needs in real-world AI applications? (RQ1)}.
Elish and Watkins \cite{elish2020repairing} recently provided insights into this question through an in-situ study of a deployed, real-world AI system. Specifically, they documented the types of inquiries physicians asked of nurses tasked with monitoring Sepsis Watch~\cite{Sendak2020SepsisWatch}, an AI system designed to predict patients' risk of sepsis development.
However, they did not study how XAI methods could answer the physicians' inquiries.
In this paper, we take a step further and contribute to understanding how XAI methods can satisfy (or not satisfy) end-users' needs by studying: \textit{How do end-users intend to use XAI explanations? (RQ2)} and \textit{How are existing XAI approaches perceived by end-users? (RQ3)}.

Our work extends prior work in three more ways.
First, while all aforementioned work~\cite{Tonekaboni2019clinicians,Cai2019CSCW,Cai2019CHI,elish2020repairing} studies AI applications that make or support high-stakes medical decisions, we focus on an ordinary application that a diverse set of people use in everyday life.
Second, while prior work does not differentiate their participants, we study group differences with respect to domain and AI background levels. We are inspired by recent findings of Ehsan and colleagues~\cite{Ehsan2021Who} on how people's perceptions of XAI explanations differed based on their AI background.
Third, we connect to the XAI methods literature directly, by mocking-up XAI explanations in the studied application. These in-situ mock-up explanations allowed us to gather detailed data on how end-users perceive and intend to use  XAI explanations in their actual use of the AI.

\subsection{XAI's role in human-AI collaboration}
\label{sec:relatedwork_collaboration}

Our work also connects to the literature of human-AI collaboration~\cite{Lai2022Content,Arous2020opencrowd,Ashktorab2020Game,Cai2019CSCW,Wang2019Datascientist,krishna2022social}, sometimes called human-AI teaming~\cite{Bansal2021Team,bansal2019hcomp,oneill2022teaming} or human-AI partnership~\cite{Nguyen2018Factchecking}, that studies how people work together with AI to achieve shared goals.
In this work, we didn't set out to study human-AI collaboration.
Our use of this term emerged from our findings: while studying participants' XAI needs, uses, and perceptions, we found that participants described a process for which the language of ``collaboration'' proved the best fit. Participants described a two-way exchange, where they help Merlin succeed in bird identification and obtain more accurate results in return, and expressed a strong desire to improve their collaboration with XAI explanations and other information.
Hence, we give a brief overview of the human-AI collaboration literature and describe how our work connects to existing work.

Prior work has studied how people collaborate with different types of AI systems (e.g., robots~\cite{Kumar2021robot,Villani2018robot,Ogenyi2021robot,Nguyen2022team,nguyen2021neurips,fel2021evaluation}, virtual agents~\cite{davella2022agent,Numata2020virtual,Ashktorab2020Game,Cila2022agent}, embedded systems~\cite{Lai2022Content,Jhaver2019Automoderator,Lai2019trust,Nguyen2018Factchecking,Amarasinghe2022frauddetection,Cai2019CSCW,Fogliato2022Clinical,Tschandl2020nature,Kim2022HIVE,Nguyen2022team,fel2021evaluation}) in different task contexts (e.g., content generation~\cite{Zhang2022StoryBuddy,Lee2022CoAuthor,Louie2020cococo}, medical diagnosis~\cite{Cai2019CSCW,Fogliato2022Clinical,Tschandl2020nature}, content moderation~\cite{Lai2022Content,Jhaver2019Automoderator}, deception detection~\cite{Lai2019trust,Nguyen2018Factchecking}, cooperative games~\cite{Ashktorab2020Game}, and fine-grained visual recognition~\cite{Kim2022HIVE,Nguyen2022team,nguyen2021neurips,fel2021evaluation}).
Among these, our work is most closely related to \cite{Lai2019trust,Kim2022HIVE,Nguyen2022team,nguyen2021neurips,fel2021evaluation} that studied XAI's role in AI-assisted decision making, where AI makes a recommendation and a human makes the final decision.
In this work, we explored what role XAI explanations could play in Merlin where for each bird identification, end-users make the final call based on the app's output and their knowledge of birds and the app.

However, different from our work, \cite{Lai2019trust,Kim2022HIVE,Nguyen2022team,nguyen2021neurips,fel2021evaluation} focused on measuring the usefulness of specific XAI methods in AI-assisted decision making through lab experiments.
These experiments typically consisted of simple tasks (e.g., binary choice) and were conducted with participants recruited from Amazon Mechanical Turk.
Further, because they were lab experiments, it was well-defined in advance how participants should use XAI explanations in their collaboration with AI (e.g., look at the provided explanation and judge whether or not to accept the AI's output).
On the other hand, our qualitative descriptive study allowed us to find that participants intended to use XAI explanations for various purposes, highlighting a broad range of XAI needs and uses that should be considered in XAI development.

\begin{figure}[t!]
\centering
\includegraphics[width=\linewidth]{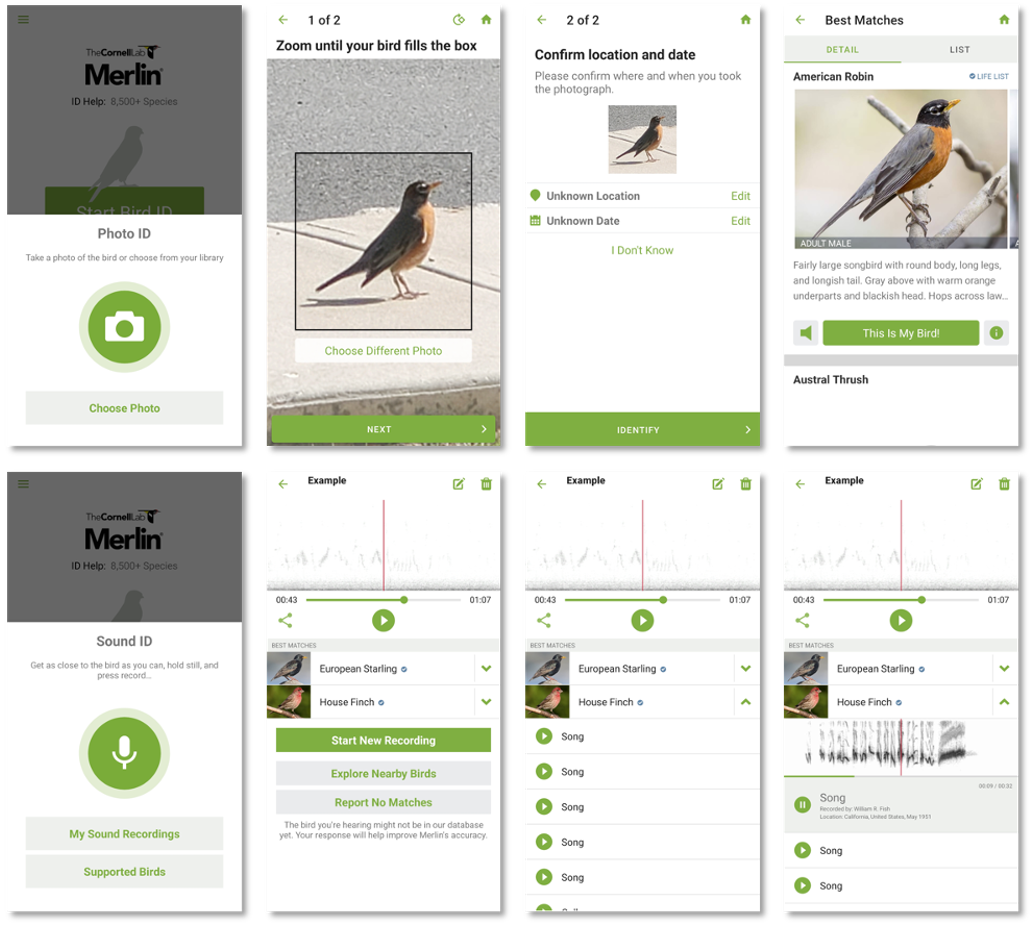}
\caption{Screenshots of Merlin, our study application. \textnormal{Merlin is an AI-based bird identification mobile phone app. Users upload photos on the \textit{Photo ID} feature (top) or sounds on the \textit{Sound ID} feature (bottom) to get a list of birds that best match the input. Users also share optional location and season data. The resulting bird list comes with example photos and sounds.}
}
\Description{This figure contains screenshots of our study application, the Merlin app for bird identification.}
\label{fig:merlin_ui}
\end{figure}

\subsection{XAI methods for computer vision}
\label{sec:relatedwork_methods}

Finally, we review the XAI methods literature to provide background on how we mocked up XAI explanations for Merlin.
We focus on methods developed for computer vision AI models because Merlin uses computer vision to identify birds in user-input photos and audio recordings. 
See \cite{arrieta2019explainable,brundage2020trustworthy,fong2020thesis,gilpin2018explaining,Gunning_Aha_2019,RudinEtAlSurvey2022,samek2019book} for more comprehensive overviews.

XAI methods can be categorized along several axes: first, whether a method is post-hoc or interpretable-by-design; second, whether it provides a global or local explanation; and third, by the explanation form.
To begin, the majority of existing XAI methods are \textit{post-hoc} methods that explain certain aspects of already-trained models~\cite{fong2019extremal,petsiuk2018rise,selvaraju2017gradcam,shitole2021sag,simonyan2013deep,zeiler2014visualizing,Zhou2016CAM,ribeiro2016lime,koh2017influence,yeh2018representer,ramaswamy2022elude,zhou2018ibd,bau2017netdissect,bau2019seeing,fong2018net2vec,kim2020gestalt}.
Recently, more \textit{interpretable-by-design} methods are being proposed; these are typically new types of computer vision models with an explicitly-interpretable reasoning process~\cite{brendel2019bagnet,Boehle2021CVPR,Boehle2022CVPR,chen2019protopnet,donnelly2022deformable,dubey2022scalable,koh2020concept,nauta2021prototree,radenovic2022neural}.
Second, XAI methods provide either a \textit{local} explanation of a model's individual output or a \textit{global} explanation of a model and its behavior.
Local, post-hoc methods include feature
attribution~\cite{fong2019extremal,petsiuk2018rise,selvaraju2017gradcam,shitole2021sag,simonyan2013deep,zeiler2014visualizing,Zhou2016CAM}, 
approximation~\cite{ribeiro2016lime}, and sample importance~\cite{koh2017influence,yeh2018representer} methods.
Global, post-hoc methods include methods that generate class-level explanations~\cite{ramaswamy2022elude,zhou2018ibd} and summaries of what a model has learned~\cite{bau2017netdissect,bau2019seeing,fong2018net2vec,kim2020gestalt}.
Interpretable-by-design models can provide local and/or global explanations, depending on the model type.
Lastly, explanations come in a variety of forms.
Representative ones are \textit{heatmaps}~\cite{fong2017meaningful,petsiuk2018rise,selvaraju2017gradcam,shitole2021sag,simonyan2013deep,zeiler2014visualizing,Zhou2016CAM,brendel2019bagnet,Wang_2020_CVPR}, \textit{examples}~\cite{koh2017influence,yeh2018representer}, \textit{concepts}~\cite{ramaswamy2022elude,zhou2018ibd,koh2020concept}, and \textit{prototypes}~\cite{chen2019protopnet,donnelly2022deformable,Nguyen2022team,nauta2021prototree}.
To the best of our knowledge, these cover the range of XAI methods for computer vision.

Since we are not affiliated with the Merlin development team and do not have access to its AI models, it was not possible to produce \textit{actual} explanations of how Merlin identifies birds. Hence, we created \textit{mock-up} explanations. For comprehensiveness, we mocked up all four aforementioned explanation forms. We know they all are plausible XAI approaches for Merlin because they have been demonstrated on bird image classification models in prior work (e.g., heatmaps in \cite{Kim2022HIVE,Wang_2020_CVPR,Pazzani2022Expert}, examples in \cite{Nguyen2022team}, concepts in \cite{koh2020concept,ramaswamy2022elude,ramaswamy2022overlookedfactors}, prototypes in \cite{Nguyen2022team,chen2019protopnet,nauta2021prototree,donnelly2022deformable}). See Fig. \ref{fig:xai_eg} and Sec. \ref{sec:interviewinstrument} for the mock-ups and their descriptions, and the supplementary material for details about how we created the mock-ups.

\section{Study application: Merlin bird identification app}
\label{sec:researchcontext}

As described in Sec. \ref{sec:relatedwork}, we looked for a research setting that involves real-world AI use by end-users with a diverse domain and AI knowledge base, and that people use in ordinary, everyday life scenarios. Furthermore, we looked for a domain with significant AI and XAI research.
We found Merlin~\cite{merlin} fit what we were looking for. Merlin is a mobile phone app (Fig. \ref{fig:merlin_ui}) with over a million downloads that end-users, with diverse birding and AI knowledge, use for bird identification as they go out and about outdoors. 
Most birding apps are digital field guides that don't use AI (e.g., Audubon Bird Guide~\cite{audubon}, iBird Pro Guide~\cite{ibird}, Birdadvisor 360\textdegree~\cite{birdadvisor}). Merlin is unique in that it uses computer vision AI models to identify birds in user-input photos and audio recordings. 

Merlin provided a grounded context with real end-users whose experience we can augment with mock-ups of XAI explanations. Furthermore, a large proportion of XAI methods for computer vision have been developed and evaluated on bird image classification~\cite{nauta2021prototree,chen2019protopnet,koh2020concept,goyal2019counterfactual,vandenhende2022counterfactual,dubey2022scalable,radenovic2022neural,Nguyen2022team,donnelly2022deformable,Pazzani2022Expert} using the Caltech-UCSD Birds (CUB) dataset~\cite{WahCUB_200_2011}. Hence, the feedback we collect on the mock-up explanations for Merlin can provide concrete and immediate insights to XAI researchers.

\section{Methods}
\label{sec:methods}

In this section, we describe our study methods, all of which were reviewed and approved by our Institutional Review Board prior to conducting the study.

\begin{table}[t!]
    \centering
    \caption{Participants' domain (bird) and AI background. \textnormal{See Sec. \ref{sec:participantbackground} for a description of the background levels.}}
    \Description{This table shows individual participants' domain and AI background. The same information is presented in Sec. 4.1.}
    \label{tab:participant}
    \begin{tabular}{llll}
    \toprule
    & \textbf{Low-AI} & \textbf{Medium-AI} & \textbf{High-AI} \\
    \midrule
    \textbf{Low-domain} & P7, P12, P16 & P8, P14 & P11, P13 \\
    \textbf{Medium-domain} & P2, P20 & P1, P4, P10 & P6 \\
    \textbf{High-domain} & P5, P17 & P3, P9, P15 & P18, P19 \\
    \bottomrule
    \end{tabular}
\end{table}

\subsection{Participant recruitment and selection} 
\label{sec:participantbackground}

We recruited participants who are end-users of Merlin's Photo ID and/or Sound ID, the app's AI-based bird identification features, with considerations for diversity in participants' domain and AI background.
Concretely, we created a screening survey with questions about the respondent's domain background, AI background, and app usage pattern (e.g., regularly used app features, frequency of app use).
We posted the survey on a variety of channels: Birding International Discord, AI for Conservation Slack, various Slack workspaces within our institution, and Twitter. On Twitter, in addition to posting the survey, we reached out to accounts with tweets about Merlin via @mentions and Direct Messages.

Based on the screening survey responses, we selectively enrolled participants to maximize the diversity of domain and AI background of the study sample. 
See Tab. \ref{tab:participant} for a summary of participants' background. The subgroups were defined based on participants' survey responses and interview answers. We refer to individual participants by identifier P\#.
\begin{itemize}[noitemsep,topsep=0pt]
    \item \textit{Low-domain}: From ``don't know anything about birds'' (P11, P12) to ``recently started birding'' (P7, P8, P13, P14, P16). Participants who selected the latter option typically have been birding for a few months or more than a year but in an on-and-off way, and were able to identify some local birds.
    \item \textit{Medium-domain}: Have been birding for a few years and/or can identify most local birds (P1, P2, P4, P6, P10, P20).
    \item \textit{High-domain}: Have been birding for more than a few years and/or do bird-related work (e.g., ornithologist) (P3, P5, P9, P15, P17, P18, P19).
    \item \textit{Low-AI}: From ``don't know anything about AI'' (P16, P17) to ``have heard about a few AI concepts or applications'' (P2, P5, P7, P12, P20). Participants in this group either didn't know that Merlin uses AI \interview{12,16} or knew but weren't familiar with the technical aspects of AI \interview{2,5,7,17,20}.
    \item \textit{Medium-AI}: From ``know the basics of AI and can hold a short conversation about it'' (P1, P3, P8, P9, P14) to ``have taken a course in AI or have experience working with an AI system'' (P4, P10, P15). Participants in this group had a general idea of how Merlin's AI might work, e.g., it is neural network based and has learned to identify birds based on large amounts of labeled examples.
    \item \textit{High-AI}: Use, study, or work with AI in day-to-day life (P6, P11, P13, P18, P19). Participants in this group were extremely familiar with AI in general and had detailed ideas of how Merlin's AI might work at the level of specific data processing techniques, model architectures, and training algorithms.
\end{itemize}
Note that our referral here and elsewhere to ``high-AI background'' participants describes their expertise with AI in general, not necessarily with Merlin's AI.
All participants were active Merlin users who could provide vivid anecdotes of when the app worked well and not. 
Regarding frequency of use, 11 participants used it several times a week, 8 used it once a week, and one used it once a month.

\subsection{Study instrument}
\label{sec:interviewinstrument}

\begin{figure*}[h!]
\centering
\includegraphics[width=\linewidth]{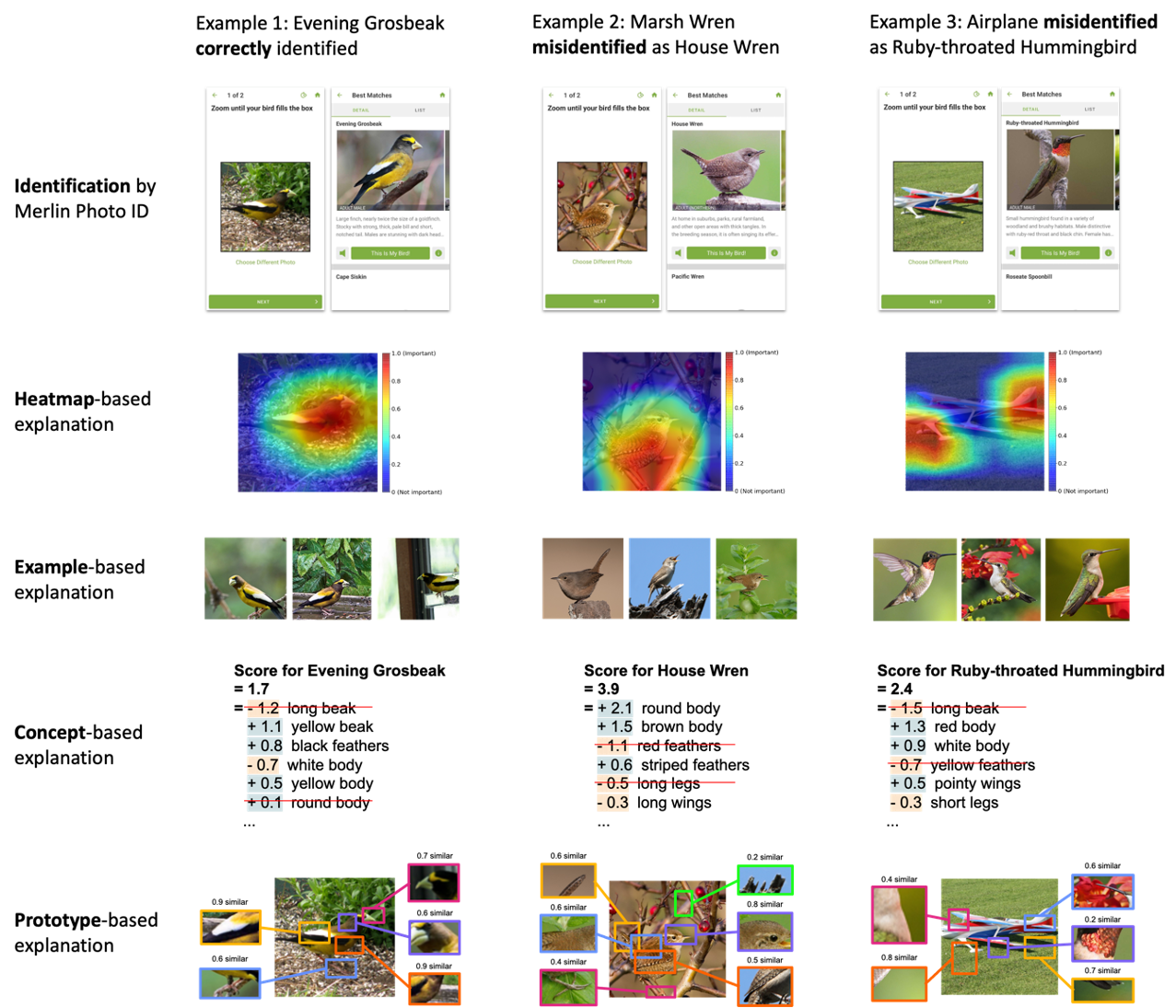}
\caption{Identification examples and XAI explanations shown to participants.
    \textnormal{
        We showed three \textit{real} examples of Merlin Photo ID identifications, paired with \textit{mock-up} explanations we designed for each identification. 
        Each XAI approach was introduced with the following description.
        \textit{Heatmap}: The AI model considers the red highlighted regions as evidence for [output, i.e., Evening Grosbeak. House Wren, Ruby-throated Hummingbird].
        \textit{Example}: The AI model considers the input photo most similar to the below photos of [output] the AI model has seen.
        \textit{Concept}: The AI model considers the presence of the following concepts as positive/negative evidence for [output].
        \textit{Prototype}: The AI model considers the boxed regions of the input photo similar to parts of [output] photos the AI model has seen.
        See Sec. \ref{sec:interviewinstrument} for details.
    }
}
\Description{This figure contains the three identification examples and the associated XAI explanations shown to participants. See Sec. 4.2 for a description.}
\label{fig:xai_eg}
\end{figure*}

Our interviews were structured in three parts and included a short survey and an interactive feedback session. 
The complete study instrument is attached in the supplementary material.
\begin{description}
    \item[Context] First, we asked the participant a series of questions aimed at learning the context of their app use. These include questions about their background; when, where, why, and how they use the app; stakes in their use; and their knowledge and perception of AI. 
    \item[XAI needs] Next, we inquired about the participant's explainability needs through open-ended questions and a survey we developed from the XAI Question Bank~\cite{Liao_CHI_2020}. The survey lists 10 categories of questions that people might have about an AI system. Nine categories (\textit{Data}, \textit{Output}, \textit{Performance}, \textit{How}, \textit{Why}, \textit{Why not}, \textit{What if}, \textit{How to be that}, \textit{How to still be this}) are from \cite{Liao_CHI_2020}, and we added a new \textit{Transparency} category on expert and social transparency~\cite{Ehsan2021ST}. The survey asks the participant to select questions they ``know the answer to'' and/or are ``curious to know (more).'' We directed the participant to interpret ``know the answer to'' as ``have a good idea of the answer'' for questions whose exact answers are not available to end-users (e.g., What is the size of the data?).
    \item[XAI uses and perceptions] Finally, we assessed the participant's perception of existing XAI approaches. Using screen sharing during the Zoom interview, we showed three examples of Merlin Photo ID identifications: the first is a correct identification; the second is a misidentification, one that people---even experienced birders---would make; and the third is a misidentification, but one that people wouldn't make. Using these as running examples, we introduced four XAI approaches one at a time and in random order (see the next paragraph for more information). For each, we asked the participant what they like and dislike about the approach, what they think can be improved, whether they want to see it in the app, and how much it helps them understand the AI's reasoning and output. See Fig. \ref{fig:xai_eg} for the identification examples and XAI explanations shown to participants.
\end{description}

To get detailed answers from participants about XAI uses and perceptions, we created mock-ups of representative XAI approaches that could potentially be embedded into Merlin.
These included: 
\begin{itemize}
    \item \textit{Heatmap}-based explanations that highlight regions in the input image that are important for the AI model’s output. 
    They represent feature attribution methods that visualize results via heatmaps (also known as saliency maps)~\cite{fong2017meaningful,petsiuk2018rise,selvaraju2017gradcam,shitole2021sag,simonyan2013deep,zeiler2014visualizing,Zhou2016CAM,brendel2019bagnet} and include popular techniques like Grad-CAM~\cite{selvaraju2017gradcam}.
    \item \textit{Example}-based explanations that show examples in the training data that are important for the AI model’s output. They include methods that use influence functions~\cite{koh2017influence} and representer points~\cite{yeh2018representer} to identify important positive/negative training examples for a particular output.
    \item \textit{Concept}-based explanations that explain the AI model's output with text-based concepts. They include concept bottleneck models~\cite{koh2020concept}, as well as methods like IBD~\cite{zhou2018ibd} and ELUDE~\cite{ramaswamy2022elude} that generate class-level explanations as a linear combination of concepts.
    \item \textit{Prototype}-based explanations that explain the AI model's output with visual prototypical parts. They represent methods such as ProtoPNet~\cite{chen2019protopnet}, ProtoTree~\cite{nauta2021prototree}, and their recent variations~\cite{donnelly2022deformable,Nguyen2022team}.
\end{itemize}
As described in Sec. \ref{sec:relatedwork_methods}, these cover the range of XAI methods for computer vision.
In the supplementary material, we describe in detail how we created the mock-up explanations and show examples of explanations beyond bird image classification. 
We emphasize that the three identification examples are \textit{real} app outputs that we obtained by uploading photos to Merlin Photo ID. However, the shown XAI explanations are \textit{mock-ups} that we designed; they are not representative of how Merlin Photo ID actually identifies birds. We made this point clear to participants during the interviews.
We also communicated that we were not affiliated with Merlin's AI development team.

\subsection{Conducting and analyzing interviews} 

We interviewed 20 participants, each over a Zoom video call, from July to August 2022. The interviews lasted one hour on average.
Participants were compensated with their choice of a 30 USD gift card or a donation to a bird conservation organization made on their behalf. 
We transcribed the interviews and then analyzed the transcripts.
First, two authors read through five transcripts to develop an initial shared codebook, primarily using descriptive coding, i.e., describing participants' expressions of their beliefs, perceptions, and actions. Descriptions were at the semantic level within the surface meanings of the data (e.g., \textit{desire to learn from the AI to improve their task skills}). Next, during multiple group meetings, all authors iterated on and refined the codebook, by shifting from descriptions of participants' beliefs to identification of shared latent, conceptual themes \cite{saldana2021coding}. \textit{Desire for improved human-AI collaboration} is an example of a latent, conceptual theme we interpreted based on participants' expressed needs for information that would help them understand the AI's capabilities and limitations, identify sources of errors, and supply better inputs, in order to more effectively interact with the AI and achieve better outcomes. After we collectively agreed that our conceptual themes were exhaustive, we then identified and eliminated themes which were redundant or overlapping. Once we had a final revised codebook, one author then used this codebook to re-code all of the data. Example codes include: \textit{desire to know the AI's confidence in its output} (XAI needs), \textit{learn how to take better pictures/audio recordings for the AI} (XAI uses), and \textit{heatmap-based explanations are too coarse} (XAI perceptions). We deliberately did not calculate inter-rater reliability (IRR) as part of our analytic process. McDonald and colleagues \cite{mcdonald2019reliability} argue that such a calculative process is a poor fit for the interpretive paradigm from which qualitative research has developed. Our codebook is derived from our collective and agreed-upon interpretations of our participants' responses to our questions, and so a mathematical post-hoc comparison of individuals' coding selections would bring little rigor to the process. Instead, we focused on bringing rigor to our analysis through the discussions and selections involved in the codebook development.

\section{Results}
\label{sec:results}

We present our results in three parts. We begin by describing participants' explainability needs (RQ1, Sec. \ref{sec:xai_needs}). We then describe how they intended to use XAI explanations (RQ2, Sec. \ref{sec:xai_use}). Finally, we describe how they perceived the four representative XAI approaches we mocked-up for Merlin (RQ3, Sec. \ref{sec:xai_feedback}).

\subsection{XAI needs: Participants desired more information about AI, especially practically useful information that can improve their collaboration with AI}
\label{sec:xai_needs}

Based on open-ended questions and the survey we developed from the XAI Question Bank~\cite{Liao_CHI_2020}, we found that while participants were generally curious about AI system details, only those with high-AI background and/or high-domain interest were willing to actively seek out this information (Sec. \ref{sec:need_difference}).
However, participants unanimously expressed a need for information that can improve their collaboration with the AI system (Sec. \ref{sec:practicalinfo}).

\subsubsection{Participants were generally curious about AI system details, but curiosity levels differed based on AI background and domain interest}
\label{sec:need_difference}

As most other AI applications, Merlin does not provide much information about its underlying technology. Hence, when we asked participants what they knew about the app's AI, all replied that they didn't know much about system details, although those with high-AI background \interview{6,11,13,18,19} had detailed guesses about the app's data, model architectures, and training algorithms.

So what did participants want to know?
According to the survey results, participants wanted to know everything about the app's AI.
For all questions in the survey, most if not all participants responded they ``know (or have a good idea of) the answer'' and/or are ``curious to know (more).'' 
That is, participants were curious about overall system details (questions in the \textit{Data}, \textit{Output}, \textit{Performance}, \textit{How}, \textit{Transparency} categories), as well as how the AI reasons and makes judgments on specific inputs (questions in the \textit{Why}, \textit{Why not}, \textit{What if}, \textit{How to be that}, \textit{How to still be this} categories).
We report the full survey results in the supplementary material.

But how curious are they, really? When we tempered self-reported levels of curiosity with interview questions about the effort participants were willing to invest to satisfy that curiosity, the picture changed. \shortquote{I wouldn't go tremendously out of my way to find the answer to these questions} \interview{12} was a sentiment shared by many participants \interview{1,5,6,7,9,10,12,13,16,20}.
For instance, P5 said: 
\shortquote{If there's an opportunity that arises, I'd love to ask about it [...] but I don't think I would be contacting people at Cornell [app developers].}
Other participants were open to searching around a bit \interview{9,10}, listening to talks or podcasts \interview{12}, or reading some documentation if easily available \interview{1,6,7,13,16,20}, but didn't want to take the initiative to seek out more information about the AI system, as described by the questions in the survey.

Exceptions were some participants with \textit{high-AI background} \interview{11,18,19} or notably \textit{high interest in birds} \interview{1,4,8}.
P11, P18, and P19, likely because they develop AI systems in their work, were very curious about the app's AI and were willing to go to the extent of reaching out to the app developers \interview{11,18} or playing with the data themselves \interview{19}.
For example, P19 said: \shortquote{I'd love to talk to one of the engineers and pick their brain [...] or get some data and play with it myself.}
P1, P4, P8 have medium-AI background, but their exceptionally high interest in birds seemed to fuel their curiosity about the app's AI. They were particularly curious about how the AI tackles difficult identifications such as mockingbirds that mimic other birds or birds that are difficult for experienced human birders to identify (e.g., ``little brown birds'').

In contrast, participants with \textit{low-to-medium AI background} \interview{7,8,9,10,12,16} had lower explainability needs.
For instance, P7, P8, and P10 had little-to-no interest in how the AI reasons and makes judgments on specific inputs.
P8 said questions in the \textit{Why}, \textit{Why not}, \textit{What if}, \textit{How to be that}, \textit{How to still be this} categories were not what they would ever think about on their own.
P7 expressed more bluntly that they prefer to keep the AI as a black box: \shortquote{No, I don't want to ruin the mystique.}
P9, P12, and P16, on the other hand, became more curious during the interview, however, their responses suggest that they were not very curious about the AI in their natural use environment prior to the interview.

In short, all participants were interested in learning more about the AI, but only those with high-AI background and/or high-domain interest were willing to expend effort to gain more information about the AI's system details.

\subsubsection{Participants desired information that can improve their collaboration with AI}
\label{sec:practicalinfo}

Participants' expressed needs for explanation shifted, however, when our interview questions moved away from gauging their curiosity about AI system details, and towards querying their use of the app. While participants' needs for system details differed based on background and interest, they unanimously expressed a need for practically useful information that could improve their collaboration with the AI system.

To begin, participants wanted a general understanding of the AI's capabilities and limitations \interview{1,4,5,16,19,20}.
P1 described a number of ways this understanding would help their use of the app:
\shortquote{It would definitely first help me understand more about when certain identifications may be more or less reliable. But also it will help me supply better inputs to the app to try and get the best quality identification results that I can} \interview{1}.
Participants had already tried to gain this understanding by pushing the AI to its limits \interview{4,5,16,19,20}.
Some had tried to fool the AI with non-bird sounds (e.g., sounds of other animals, bird impersonations) to understand when it works and when it breaks \interview{4,5,16,19}.
Others had conducted more rigorous experimentation by altering their input (e.g., clip the audio recording, remove location information) and observing changes in the AI's output to understand what factors influence the AI's output and how \interview{4,20}.

Another frequently expressed need was for a display of the AI's confidence \interview{1,2,3,4,6,13,18,20}. Participants wanted this information to better determine when to trust the AI's output.
Concretely, P2 demanded for percentage-based confidence scores: 
\shortquote{If it doesn't give a percentage [...] I just don't have a gauge of how correct it is} \interview{2}.
P7 requested the AI to qualify its output by saying ``it may not be the exact match'' or give a general answer (e.g., ``we don't know the exact species but this bird is in the Wren family'').

Lastly, participants wanted the AI to give more \textit{detailed outputs} \interview{2,10,11,12}.
They demanded information that would help them verify the AI's output.
For instance, P10 wanted the AI to \shortquote{highlight the time period of the [sound] clip that it calls a certain species} because it is hard to know which sound corresponds to which bird when multiple birds are singing at once.
Going a step further, P2, P11, and P12 wanted the AI to 
specify the type of bird sound it heard.
Currently, the verification process is arduous because each bird species has a number of songs and calls, as well more specific sounds such as juvenile calls, flock calls, and alarm calls.
They said the suggested features will make the verification process easier and provide more information about how the AI has made its identification, with which they can more readily check the AI's output and determine whether to trust it. 

In sum, when we queried participants about their actual, real-world use of the app, they expressed a desire for information which could improve their use of the app, particularly in deciding whether or not to trust the AI's outputs. Intriguingly, they expressed these desires before we showed them our mock-ups of what XAI explanations for the app might look like.
This suggests that these XAI needs were not prompted solely by seeing XAI explanations.

\subsection{XAI uses: Participants intended to use explanations for calibrating trust, improving their own task skills, collaborating more effectively with AI, and giving constructive feedback to developers}
\label{sec:xai_use}

Next, when we showed XAI explanations to participants, they were excited to use them for various purposes beyond understanding
the AI’s outputs: for determining when to trust the AI (Sec. \ref{sec:use_trust}), which is a well-known use and commonly-stated motivation for XAI \cite{Zhang2020FAccT,Yin2019trust,Scharowski2022TRAIT,Miller2022TRAIT,Ferrario2022FAccT}, but also for learning to perform the task better on their own (Sec. \ref{sec:use_learn}), changing their behavior to supply better inputs to the AI (Sec. \ref{sec:use_action}), and giving feedback to the developers to improve the AI (Sec. \ref{sec:use_feedback}), which are less discussed uses in existing literature.

\subsubsection{Participants intended to use explanations to determine when to trust AI}
\label{sec:use_trust}

Many participants said they would use explanations to determine when to believe the app's identification result \interview{1,4,8,11,13,18,20}.
The need underlying this use is consistent with the aforementioned need for information that helps them decide when to trust the AI.
While looking at different explanation mock-ups, participants gave examples of when their trust would increase and decrease.
For instance, participants said they would feel more confident in the AI's output when heatmap-based explanations show that the AI is \shortquote{looking at the right things} \interview{8} and when example-based explanations show example photos that look similar to their input photo.
Conversely, they said they would feel more skeptical when heatmap-based explanations suggest that an \shortquote{artifact was important} \interview{8},
when concept-based explanations have errors in their concept recognition (e.g., says there is a long beak when there is not) \interview{18},
and when prototype-based explanations match photo regions and prototypes that don't look similar to them \interview{4}.
These findings confirm existing literature~\cite{Zhang2020FAccT,Yin2019trust,Scharowski2022TRAIT,Miller2022TRAIT,Ferrario2022FAccT} and suggest that trust calibration will be an important use of XAI.

\subsubsection{Participants desired to learn via explanations to better achieve the task on their own}
\label{sec:use_learn}

Intriguingly, a greater number of participants said that they intend to use explanations to improve their task skills \interview{1,2,4,6,7,8,9,10,11,13,15,17,19,20}.
Participants viewed the AI as a teacher and were keen to learn the features it looks at via explanations, so they can look for these features in the future when they are birding on their own.
Participants were aware that the features the AI looks at may be different from what expert human birders look at. 
But they weren't very concerned about the potential differences.
One participant even said it would be interesting if the AI finds new ways of identifying birds and explanations can \shortquote{call attention towards things that people did not really think of before} \interview{1}. Still, participants preferred that explanation \textit{forms} be similar to those of human birders.
We elaborate on this point further in Sec. \ref{sec:xai_feedback}.

Overall, participants were excited about how explanations could make birding more accessible for themselves and others who lack access to expert resources (e.g., mentoring from human birders):
\longquote{It [the explanation] is kind of training or giving me more information and I'm kind of learning these things [what features to look at]. Whereas before, birders or ornithologists are learning this from mentors or teachers in the field. But those opportunities are limited based on social relations, privilege, how closely you are are connected to birding groups and stuff. And so it will be much more openly accessible if that kind of more comparative identification knowledge was accessible through just an app.}{1}
Even participants with high-domain background, whose main goal for using the app was not to obtain such knowledge, appreciated the educational value of explanations and said explanations would help them learn faster \interview{16}.

These findings are closely related to recent works by Goyal and colleagues~\cite{goyal2019counterfactual} and Pazzani and colleagues~\cite{Pazzani2022Expert}. They demonstrated that XAI explanations help non-bird-experts (graduate students in machine learning \cite{goyal2019counterfactual} and undergraduate students in psychology, cognitive science, or linguistics courses \cite{Pazzani2022Expert}) learn to distinguish birds. While their experiments employed relatively easy tasks, i.e., assigning bird images to one of two species options, they showed the potential of \textit{learning from AI via XAI explanations}. While \cite{goyal2019counterfactual,Pazzani2022Expert} did not establish that this is a need that people have, our work provides empirical evidence for it, suggesting \textit{learning from AI} as another important use case for XAI.

We postulate this use case stemmed from Merlin's status as an expert AI system.
Many AI applications are deployed to automate tasks that are easy for people (e.g., face verification, customer service chatbot) in settings where it is costly or implausible to have humans in the loop. In contrast, Merlin possesses expertise that most people don't have and need to invest time and effort to gain.
This expertise is likely the source of Merlin explanations' educational value. In other types of AI applications, end-users may not intend to learn from AI via explanations.

\subsubsection{Participants viewed explanations as an opportunity to be better AI-collaborators}
\label{sec:use_action}

Participants also saw explanations as an opportunity for action. They looked for feedback on their own behavior that would in turn enable them to help the AI better achieve the task \interview{1,7,9,20}.
P20 said explanations, by providing insights into how the AI got an identification wrong, can help them figure out the answer to: \shortquote{What would I have to do to change this photo to make it [AI] understand it better?}
Participants sought out opportunities to improve their own collaborative skills when working with the AI to achieve a task, because at the end of the day they want to achieve best possible outcomes:
\longquote{You're still trying to look for the right bird. So if you can adjust human behavior to get the right answer out of the robot [AI], then that's helpful.}{20}
Because of this need, participants were critical towards XAI approaches they thought didn't provide actionable feedback.
For instance, P9 questioned the utility of heatmap and example-based explanations: 
\shortquote{How is it helpful to the user in the future? Besides just being cool and interesting? How does it change the user's use of the app? Does it make you take a different photo?}
They critiqued that these approaches don't help them help the AI be more correct.

We view use this intended use of XAI explanations as an extension of participants' current efforts to help out the AI.
When describing their use of the app, participants mentioned several different ways they help the AI perform better.
Some were smaller adjustments on the spot, such as facing the microphone closer to the bird and getting a sufficiently long recording for Sound ID \interview{9}.
Others were more involved, such as the efforts P1 described as part of their \shortquote{general workflow} for using Photo ID:
\longquote{I basically don't use images that are either too blurry or do not feature the bird in an unobstructed manner. I know from my personal experience using it that Merlin works a lot better if it has a more silhouetted side profile shot of the bird. [...] So I try to feed Merlin photos taken from similar angles, also in acceptable lighting conditions. I might have to boost the contrast or the brightness of a picture artificially to feed it into Merlin to get better results. If there's no real contrast, then it's much harder to get credible results.}{1}

In short, participants viewed the AI as a collaborator.
They have already found ways to better work with it, 
and they intended to use XAI explanations to further improve their collaboration. 
To this end, they wanted explanations to provide actionable feedback on their own behavior so that they can supply better inputs to the AI.

\subsubsection{Participants saw explanations as a medium to give feedback to developers and improve AI}
\label{sec:use_feedback}

Finally, participants with high-AI background intended to use explanations as a medium to give feedback to developers and contribute to improving the AI \interview{13,18,19}.
These participants mentioned that explanations, by providing more information to end-users about how the AI produced its output, enable end-users to give more detailed feedback. This feedback can then help developers improve the AI system.
P13 illustrated this process using prototype-based explanations as an example:
\longquote{The fact that it [AI] identifies parts of the tree, that's a great opportunity to [to have end-users] tap that region and say `not a part of the bird' so that you can get the users helping you to do some curation and labeling on the images, which someone could review or whatever. You can make much higher quality models by getting this sort of the labeling right.}{13}
P18 suggested a similar feedback process for example-based explanations. They said when end-users disagree with the provided examples of similar looking birds, they can correct them by saying \shortquote{no, I think it actually looks more like bird number three} and help developers align the AI's notion of perceptual similarity with that of humans,
and improve the AI.

Lastly, P19 described XAI's potential for creating a positive feedback loop that helps both end-users and the AI system:
\longquote{So there's a feedback loop here, right? Because if that [teaching people to better identify birds] is your goal, and you're successful in doing that, then you're able to rely on people to verify their data, contribute solid data, and that data can help inform Merlin, which makes Merlin better, which makes it do its job better. [...] I think no matter what, it [providing explanations] is kind of beneficial.}{19}
P13 and P18 shared this view and said they would be excited to help developers improve the app by providing feedback via explanations. P18, in particular, expressed a strong desire to contribute. They had already been signing up for beta versions of the app, and the first answer they gave to the question ``What would you like to know more about Merlin?'' was: \shortquote{How I can contribute more} \interview{18}.

In short, participants with high-AI background desired to use explanations to help improve the AI, so that they can achieve better outcomes with it in the future. We interpret this as another example of participants viewing the AI as a collaborator whom they work together with.

\subsection{XAI perceptions: Participants preferred part-based explanations that resemble human reasoning and explanations}
\label{sec:xai_feedback}

In this last results section, we describe how participants perceived the four XAI approaches we mocked up: Heatmap (Sec. \ref{sec:feedback_heatmap}), Example (Sec. \ref{sec:feedback_example}), Concept (Sec. \ref{sec:feedback_concept}), and Prototype (Sec. \ref{sec:feedback_prototype}). We also summarize concerns expressed toward explanations (Sec. \ref{sec:feedback_concerns}), and explore how existing XAI approaches might satisfy end-users' explainability needs and goals identified in the previous sections.

\subsubsection{Heatmap-based explanations: Most mixed opinions}
\label{sec:feedback_heatmap}

We received the most mixed reviews for heatmap-based explanations.
Participants who liked heatmaps described them as \shortquote{fun} \interview{15}, \shortquote{aesthetically pleasing} \interview{3}, and intuitive---\shortquote{it's very easy, it hits you right away} \interview{9}.
Some participants were positive because they often use heatmaps in their work and find them helpful for representing information \interview{12,19}.
Conversely, a few participants expressed a strong dislike \interview{14,16},
e.g., \shortquote{I hate those things [...] They are simply not intuitive} \interview{14}.
P20 didn't like heatmaps as an explanation form because \shortquote{heatmaps feel like they should be related to weather,} revealing individual differences in perception.

Regarding utility, some said heatmaps help them understand how the AI had made a mistake \interview{7,9,13}. For instance, P19 said they see how the AI made a mistake for the Marsh Wren photo because the heatmap (in Fig. \ref{fig:xai_eg}) did not highlight areas that are important for distinguishing different species of Wrens (e.g., Marsh Wren has a white eyebrow that House Wren doesn't).
However, many participants criticized that the shown heatmaps were too coarse and uninformative \interview{1,2,3,4,6,10,11,16,17,19}. 
\shortquote{It's just highlighting the bird} was a common remark.
Participants said heatmaps would be more helpful if they highlight a few salient features of the bird, just like how human birders focus on a few field markers when identifying birds.

Finally, some participants thought heatmap-based explanations were inherently limited by its form.
P9, P11, and P17 said heatmaps were unsatisfying because they don't answer the ``why'' question. Regarding heatmaps' highlighted regions, P17 asked: \shortquote{Yes it's important, but why was it important?}
Other participants were dissatisfied because heatmaps lacked actionable information \interview{9,11}. They said knowing which parts of the photo were important to the AI does not help them change their behavior to help the AI be more correct in future uses.

\begin{table*}%
    \centering
    \caption{Summary of participants' feedback on four XAI approaches.
    \textnormal{See Sec. \ref{sec:xai_feedback} for details.}
    }
    \Description{This table summarizes participants' feedback on four XAI approaches. See Sec. 5.3. for details.}
    \label{tab:xai_summary}
    \begin{tabular}{p{0.12\textwidth}p{0.41\textwidth}p{0.41\textwidth}}
    \toprule
    \textbf{XAI approach} & \textbf{Praises} & \textbf{Complaints} \\
    \midrule
    \textit{Heatmap} & Intuitive, pleasing & Unintuitive, confusing \\
     & Helpful for spotting AI's mistakes & Uninformative, too coarse \\
     &  & Doesn't explain why certain parts are important \\
     &  & Doesn't give actionable information \\
    \midrule
    \textit{Example} & Intuitive & Uninformative, impression-based \\ 
     & Helpful for verifying AI's outputs & Doesn't add much to current examples \\ 
     & Allows end-users to do their own moderation & Doesn't give actionable information \\ 
    \midrule
    \textit{Concept} & Part-based form & Current concepts are too generic \\ 
     & Resembles human reasoning and explanations & Meaning of coefficients is unclear \\ 
     & Helpful for verifying AI's outputs & Numbers are overwhelming \\ 
     & Helpful for learning bird identification &  \\ 
     & Final scores and coefficients are helpful &  \\ 
    \midrule
    \textit{Prototype} & Part-based form & Cluttered, difficult to see on small screens \\ 
     & Resembles human reasoning and explanations & Some prototypes are ambiguous and uninteresting  \\ 
     & Intuitive, visual &  \\ 
     & Helpful for verifying AI's outputs &  \\ 
     & Helpful for learning bird identification & \\
    \bottomrule
    \end{tabular}
\end{table*}

\subsubsection{Example-based explanations: Intuitive but uninformative}
\label{sec:feedback_example}

There was a consensus among participants that example-based explanations are \shortquote{really easily understandable.} However, opinions diverged regarding their utility.
Some found them helpful for determining when to trust the AI \interview{4,5,17,20} since they themselves can compare their input photo to the examples in the explanations.
P4 noted that example-based explanations feel \shortquote{much more collaborative} since they allow end-users to do their own moderation of the provided information.
P19, on the other hand, were concerned that they would \shortquote{open the door for user error.}
Especially for difficult identifications where there are only subtle differences between candidate birds, P19 said example-based explanations wouldn't help non-bird-expert end-users arrive at a more accurate identification.

Many participants described example-based explanations as rather uninformative \interview{1,4,6,8,10,11,12,18}.
Some thought they didn't add much information to example photos that are already shown in the app with the identification result \interview{1,6,10,11}. They understood the difference between the two, that example-based explanations convey what the AI considers similar to the input photo, while the currently provided example photos are part of a fixed bird description and independent of the input. Still, they thought the explanations were not very useful. 
Some even preferred the current example photos because they are high-quality and well-curated \interview{1,6}.

Another frequent criticism against example-based explanations was that they are too general and impression-based \interview{4,8,10,12,18}. 
Participants were frustrated that they don't communicate what features the AI was using to make its identifications, e.g., P8 said \shortquote{This kind of tells you nothing.}
Due to this lack of specificity, many mentioned that example-based explanations were not helpful for their various intended uses, ranging from understanding the AI's reasoning to supplying better inputs to the AI to improving their own bird identification skills.

\subsubsection{Concept-based explanations: Well-liked overall but overwhelming to some}
\label{sec:feedback_concept}

Participants were largely positive towards concept-based explanations.
Most praises were about their part-based form.
They liked that the AI's output was broken down into chunks that human birders reason with, i.e., concepts \interview{3,4,11}.
\shortquote{This is what a person looks for basically when they're identifying a bird,} remarked P3.
Relatedly, participants liked that concept-based explanations resemble the way bird identifications are taught and shared between birders \interview{3,8,17}.
P17 said, \shortquote{before all this technology, this is exactly how you would basically learn to ID a bird.} 
For these reasons, participants mentioned that concept-based explanations seem helpful for learning to identify birds on their own.

Participants also mentioned other use cases where concept-based explanations can help.
For instance, P11 said they 
would
allow people to check the AI's output more thoroughly because people can agree or disagree with the explanation at the level of individual concepts.
As an example, they said they would not believe the AI's output if the explanation says there are red feathers in the photo when there are not.
Participants also liked that the shown explanations provided a final score for the output because they display the AI's confidence in the identification \interview{1,5,17}. 
P5 said such scores would be particularly helpful when they are comparing similar-looking candidate birds. 

Nonetheless, participants mentioned a few areas of improvement.
Several participants pointed out that the concepts in the shown explanations (e.g., long beak, black feathers, white body) were too general \interview{1,4,5,10}. 
They suggested adopting birders' language and describing birds with more specific terms such as \shortquote{underbelly, chest, rump, wing, wingbars, neck, head, cap} \interview{4}.
Participants also recommended making the numbers in the explanations as easily understandable as possible \interview{6,9,12,13,15,16,18}. 
P6 pointed out that the current concept coefficients are confusing: \shortquote{I have no idea what any of the numbers mean? Like is 1.7 good?}
Specifying what are good and bad numbers and constraining the coefficients' range may mitigate some of the confusions.
Even with these changes, however, concept-based explanations may not be everyone's cup of tea.
Some participants shared that they find the explanation form inherently overwhelming and less attractive \interview{5,13,16,20}.
P16 shared: \shortquote{I sort of tune out with numbers after a while.}
P20 also expressed their preferences for more visual explanations: \shortquote{I'm such a visual person that stuff like this would go right over my head and make no sense for the most part.}

\subsubsection{Prototype-based explanations: Most preferred}
\label{sec:feedback_prototype}

Many participants picked prototype-based explanations as their favorite \interview{2,3,4,6,7,9,10,12,13,15,16,17,19,20}.
The part-based form was clearly preferred, for similar reasons mentioned for concept-based explanations.
P15 and P20 said prototype-based explanations are analogous to how they think about birds, and P1 that they are analogous to how birders teach each other.
Between prototypes and concepts, participants tended to prefer prototypes for their visual nature and information content: prototype-based explanations locate and draw a box around relevant bird parts in the user-input photo, whereas concept-based explanations only list the bird parts.
P4 summarized the advantages: \shortquote{It makes a very clear match between the photo that you're looking at and a larger base of what this bird should look like. It also skips over the whole language issue and is incredibly visual which I really appreciate.}
Participants also noted that prototype-based explanations can help many uses, e.g., learning how to identify new birds \interview{2,8,13,15,19,20}, 
understanding how the AI is working \interview{11,13,15,16,20}, spotting the AI's mistakes \interview{4,13}, and changing their own behavior to supply better inputs to the AI \interview{20}.

Complaints against prototype-based explanations were mostly minor.
Some participants described the current version as \shortquote{cluttered} and \shortquote{difficult to see} \interview{1,4,5,6,11} and made UI design recommendations, e.g., having one prototype-photo region match pop up at a time \interview{11}.
Participants also mentioned that some prototypes were ambiguous \interview{2,11,18}.
For instance, P11 said they had to \shortquote{examine the prototype and the example to figure out what the concept was that they corresponded to.} As a solution, P2 suggested providing a textual description of the prototype. Another complaint was that some prototypes (e.g., feet) were uninteresting \interview{1,13,18}. \shortquote{Very few bird species are differentiated based on their feet,} remarked P1. For solving this problem, participants suggested curating prototypes with domain experts and end-users so that the explanation focuses on salient and interesting features, those that human birders would use to identify birds.

Finally, several participants suggested combining prototype-based explanations with other approaches \interview{2,4,11,12,16,18,19}. Concretely, P2 suggested combining it with heatmap-based, P2, P12, P16 and P18 with concept-based, and P4 and P11 with example-based explanations. P19 didn't specify an approach. Regarding the combination, some suggestions were general (e.g., show both types of explanations) while others were more specific (e.g., add concept labels to prototypes). P12 and P18 particularly advocated for using information from multiple sources (e.g., photo, sound, location) for both improving the AI's performance and explaining its results to end-users.

\subsubsection{Concerns about XAI explanations}
\label{sec:feedback_concerns}

Participants were overall excited to see XAI explanations in Merlin, however, some expressed concerns regarding the faithfulness and potential negative effects of explanations. In particular, participants who were familiar with XAI questioned how faithfully the shown approaches would explain the app's identification process, if they were to be implemented in the app \interview{6,10}. For example, P6 said example-based explanations feel like \shortquote{cheating interpretability} unless the AI actually makes identifications using clustering or other techniques that group similar photos together. Regarding concept-based explanations, P6 and P10 asked if they imply that the AI system is \textit{interpretable-by-design} and actually reasons in two steps (first concept recognition, then bird identification), or if they were \textit{post-hoc} explanations produced by a separate ``explainer'' system. These questions highlight the importance and challenges of communicating what XAI explanations are actually showing. In some cases, explanations of XAI explanations (``meta-explanations'') may be more complex than the XAI explanations themselves.

Another concern was that explanations might lead to mistrust or overtrust in AI systems. P20 said a convincing explanation for a misidentification would be \shortquote{detrimental} to end-users who are trying to learn bird identification on their own, because they might more readily believe in the misidentification and accumulate wrong knowledge. Similarly, P19 said explanations might encourage end-users to \shortquote{double down on the incorrect identification,} and even create a negative feedback loop if the AI system relies on end-users to input or verify data. These concerns are consistent with findings from recent research~\cite{poursabzi2021manipulating,Kim2022HIVE} that people tend to believe in AI outputs when given explanations for them, and raise caution against negative effects explanations might have on end-users, irrespective of XAI designers' intent.

\section{Discussion}
\label{sec:discussion}

\subsection{XAI as a medium for improving human-AI collaboration}
\label{sec:discussion_collaboration}

The most surprising finding from our study was the degree to which Merlin end-users wanted to improve their \textit{collaboration} with the AI system through XAI explanations.
Participants desired information upon which they can act. In particular, they wanted XAI explanations to \textit{help them help the AI}, e.g., by supplying better inputs to the AI and providing constructive feedback to developers. We found this an intriguing re-purposing of explanations, which are typically developed to help people understand AI's inner workings and outputs.
Collaboration is distinct from \textit{usability}.
Usability is already often discussed in the XAI literature, where XAI is presented as a means to provide meaningful information about how an AI output is reached, so that users can ``make decisions more quickly, or to increase decision quality'' \cite{Langer2021stakeholders}. 
However, our participants desired information that not only improves their decisions based on the AI's outputs (usability), but also empowers them to help the AI be more accurate in the first place and achieve better outcomes together (collaboration).

For designing XAI that supports human-AI collaboration, research in the accessibility field can be instructive.
Bigham and colleagues' work on the VizWiz system \cite{bigham2010vizwiz} combined computer vision with human labor to support blind end-users in solving daily visual problems. 
The VizWiz system allowed end-users to upload a photo of their environment and ask remote workers visual questions (e.g., where is the tomato soup can?).
In their study, the authors found that input photos presented challenges to the VizWiz system because the blind end-users provided photos which were ``often blurred, tilted, varied in scale, and improperly framed .. and susceptible to problems related to lighting'' \cite{bigham2010vizwiz}.
To overcome these challenges, the authors developed a subsystem that uses computer vision to provide end-users with prompts around lighting, centering, and distance-to-object.
These prompts helped end-users take photos that help remote workers better answer their questions. Like how VizWiz's subsystem helped end-users collaborate with VizWiz and remote workers, we argue XAI explanations can and should serve as a medium for improving end-users' collaboration with AI systems.
In addition to providing technical transparency, XAI explanations should move towards providing \textit{actionable feedback} to and from end-users, empowering end-users to have more rich and meaningful interactions with AI systems.

\subsection{XAI design for Merlin}

So what should Merlin's XAI explanations look like? In this section, we propose a design based on participants' perceptions of the four XAI approaches (Sec. \ref{sec:xai_feedback} and Tab. \ref{tab:xai_summary}). We hope this serves as a helpful example of how end-users' feedback can inform and guide the design of XAI explanations.

Participants preferred specific, part-based explanations that resemble human reasoning and explanations.
Participants repeatedly demanded that explanations highlight a few specific features that the AI uses to make its identifications.
XAI approaches that best satisfied this need were concept and prototype-based explanations that break down the AI's output with human-digestible units of information, i.e., concepts and prototypes.
Participants also appreciated how similar concept and prototype-based explanations were to the way human birders identify birds and explain their identification to others.
Heatmap and example-based explanations were generally less preferred: many participants noted that they were too coarse and impression-based to be useful.

At the same time, participants wanted explanations to be easily understandable.
\shortquote{I don't want to spend extra time trying to understand what I'm looking at} \interview{5} was a common sentiment.
However, what counts as easily understandable differed between participants.
For instance, participants with low-AI background found concept-based explanations overwhelming due to the numbers and calculations. Contrarily, participants with high-AI background found all explanations intuitive, with some wanting even more numbers and other technical details.
These findings underline the importance of our research agenda, as end-users' XAI needs, perceptions, and variations therein cannot be anticipated by studying only AI experts, which have made up a primary study population in XAI research.

Based on these observations, if we were to design XAI explanations for Merlin, we would combine prototype and concept-based.
For simplicity, we would start with a visualization of the user's input photo overlaid with boxes indicating regions matched to prototypes.
We would then give the user an option to tap on each box to get more information.
The pop-up would show the prototype matched to the boxed region, accompanied with a short description, i.e., concept name.
As some participants mentioned, prototypes may be ambiguous.
For instance, a prototype visualizing a part of a bird's wing may not be informative if end-users don’t know what is special about that part. A text description (e.g., white stripes across the folded wing) would help remove the ambiguity.
Further, we would curate the set of prototypes for each bird with end-users and domain experts.
We would exclude prototypes that are uninteresting (e.g., feet) or redundant, and limit the total number of prototypes so as to not overwhelm users.
Finally, we would optionally provide similarity scores, coefficients, and final class scores for end-users who are interested in diving into the details.

\subsection{Implications for future XAI research}

Our findings have broader implications for XAI research beyond designing explanations for Merlin. Below we discuss directions we believe XAI explanations should be improved to better serve the needs of AI system end-users.
\begin{enumerate}
    \item \textit{Explanations should be designed with end-users.} Participants often exposed blind spots in existing XAI approaches, revealing a creator-consumer gap in XAI research~\cite{Ehsan2021Who}. For example, they pointed out that the concepts used in concept-based explanations were disconnected from birders' language (Sec. \ref{sec:feedback_concept}). The shown concepts (e.g., white body, long wings) were too generic compared to birders' field mark terms (e.g., wingbar, supercilium). Participants also proposed solutions, e.g., develop the bank of concepts with end-users, and offered to contribute their experience and domain expertise. This example highlights the need for end-users' \textit{participation} in the explanation design process and calls for more participatory approaches~\cite{Muller2002PD} to XAI research.
    \item \textit{Explanations should answer ``why'' not just ``what.''} Several of our participants were unsatisfied with existing XAI approaches that only explain ``what'' features the AI system was using to produce its output, e.g., heatmap explanations that highlight ``what'' image regions were important but don't explain ``why'' those regions were important (Sec. \ref{sec:feedback_heatmap}). They expressed a desire for explanations that answer ``why'' question so that they can gain a deeper understanding of the AI's reasoning process. Explaining causal relationships in computer vision AI models is an open problem, however, it is promising that more researchers are tackling causal questions in XAI research~\cite{Moraffah2020causal}.
    \item \textit{Explanations should use multiple forms and modalities if warranted}. There is no reason for explanations to be limited to one form or modality. Participants often suggested combining two or more XAI approaches to produce more informative explanations (Sec. \ref{sec:feedback_prototype}). They also questioned why Merlin's identification features (i.e., Photo ID, Sound ID) and our explanation mock-ups were not multimodal, when human birders combine evidence from as many sources as possible (e.g., photo, sound, location) for more accurate bird identification. Expanding the design space of explanations will lead to XAI methods that better satisfy end-users' needs.
    \item \textit{Explanations should be rigorously evaluated.} Explanations sometimes have (unintended) negative effects. Recent works have revealed that explanations can engender over-trust in AI or give misleading understandings~\cite{Kim2022HIVE,Ehsan2021Who,adebayo2018neurips,hoffmann2021looks,lipton2017mythos,margeloiu2021concept,nguyen2021neurips,shen2020hcomp}. 
    Our participants were also concerned about the faithfulness and potential negative effects of explanations (Sec. \ref{sec:feedback_concerns}). To preemptively, not reactively, address these issues, it is crucial to rigorously evaluate XAI methods on both their method goals and use case goals throughout the development process.
\end{enumerate}

\section{Limitations and future work}
\label{sec:limitations}

First, as most of our interview questions and study materials pertain to the Merlin app, our findings may not generalize to other contexts.
This is an intentional trade-off made in favor of deepening our understanding of end-users' XAI needs, uses, and perceptions in a specific context.
However, our study design can aid future research on other types of AI applications and the disparate domains and contexts into which they are integrated.
Another limitation is that we had relatively few participants in some of the background subgroups; in future research we aim to have more participants from these subgroups.
Finally, we did not have access to other stakeholder groups, such as the developers and deployers of the Merlin app.
However, different stakeholders in AI systems might have different needs for XAI explanations~\cite{Langer2021stakeholders,Gerlings2021casestudy}.
We plan to pursue additional research with a more comparative design and study if and how needs differ across stakeholder groups.

\section{Conclusion}
\label{sec:conclusion}

There is a lack of empirical studies examining end-users' explainability needs and behaviors around XAI explanations in real-world contexts. This is important to address in order to make AI systems accessible, usable, and understandable for more people. 
In this work, we conducted a qualitative, descriptive, and empirical study with 20 end-users of the Merlin bird identification app. 
Our questions revolved around real-world needs and usage, with a goal to surface insights which can be utilized to improve the design of XAI explanations.
We found interesting insights into the actionability with which participants collaborate with the AI system and desire to use explanations to improve their collaboration.
We also gathered concrete feedback on four representative XAI approaches that could be potentially embedded into Merlin, finding that participants prefer part-based explanations that resemble human reasoning and explanations. Further, participants' feedback revealed a creator-consumer gap in XAI, highlighting a need of involving end-users in the XAI design process. Based on our findings, we provided recommendations for future XAI research and design.

\begin{acks}
We foremost thank our participants for generously sharing their time and experiences.
We also thank Tristen Godfrey, Dyanne Ahn, and Klea Tryfoni for their help in the interview transcription.
Finally, we thank members of the Princeton HCI Lab and the Princeton Visual AI Lab, especially Amna Liaqat and Fannie Liu, and the anonymous reviewers for their thoughtful and helpful feedback.

This material is based upon work partially supported by the National Science Foundation (NSF) under Grant No. 1763642 awarded to OR. Any opinions, findings, and conclusions or recommendations expressed in this material are those of the authors and do not necessarily reflect the views of the NSF.
We also acknowledge support from the Princeton SEAS Howard B. Wentz, Jr. Junior Faculty Award (OR), Princeton SEAS Project X Fund (RF, OR), Princeton Center for Information Technology Policy (EW), Open Philanthropy (RF, OR), and NSF Graduate Research Fellowship (SK).
\end{acks}


\bibliographystyle{ACM-Reference-Format}
\bibliography{references}


\end{document}